\documentclass[a4paper,11pt]{article}
\usepackage{jheppub} 

\newcommand {\C}{{\mathbb C}}
\newcommand {\R}{{\mathbb R}}
\newcommand {\CP}{{\mathbb {CP}}}
\usepackage{float}
\usepackage{caption}
\usepackage{subcaption}

\title{Singularities of Feynman Integrals}

\author[a]{Tanay Pathak}
\author[b]{Ramesh Sreekantan}
\affiliation[a]{Centre for High Energy Physics, Indian Institute of Science,\\ Bangalore-560012, Karnataka, India}
\affiliation[b]{Statistics and Mathematics Department,
Indian Statistical Institute, Bangalore,
Bangalore-560059. Karnataka, India
}

\emailAdd{tanaypathak@iisc.ac.in,rsreekantan@isibang.ac.in}

\abstract{In this paper, we study the singularities of Feynman integrals using homological techniques. We analyse the Feynman integrals by compactifying the integration domain as well as the ambient space by embedding them in higher-dimensional space. In this compactified space the singularities occur due to the meeting of compactified propagators at non-general position. The present analysis, which had been previously used only for the singularities of second-type, is used to study other kinds of singularities viz threshold, pseudo-threshold and anomalous threshold singularities. We study various one-loop and two-loop examples and obtain their singularities. We also present observations based on results obtained, that allow us to determine whether the singularities lie on the physical sheet or not for some simple cases. Thus this work at the frontier of our knowledge of Feynman integral calculus sheds insight into the analytic structure.}

\begin{document}
\maketitle
\flushbottom
\section{Introduction}

Feynman integrals are important for precision calculations in quantum field theory. Their study is a very mature field, with a large number of techniques, computational, numerical and analytic unifying and exemplifying several branches of mathematics. It may be worth recalling that Feynman diagrams per se are now over 7 decades old.  While many of their technical properties have been known for several decades, there are aspects that have been studied in the past using techniques of that era, which have not been sufficiently developed for one reason or another.  With the focus shifting to the Standard Model of the electro-weak and strong interactions, and to properties of field theories including that of renormalization and of renormalization group, and with the advent of dimensional regularization as the favoured method for regularization in most instances, a large number of results are today available at higher loops and with several masses and with several external legs.  They can be evaluated using many techniques, Mellin-Barnes techniques and differential equation techniques, to name a couple\cite{smirnov2006feynman,Weinzierl:2022eaz,Ananthanarayan:2020ncn,Ananthanarayan:2020fhl}. Apart from their evaluation they also have a rich mathematical structure for example, a Hopf algebra structure\cite{Abreu:2014cla,Abreu:2015zaa,Abreu:2017ptx,Abreu:2017mtm,Ananthanarayan:2021cch}, coaction \cite{Abreu:2017enx}, twisted cosmology groups\cite{Abreu:2022mfk}, homology group \cite{federbush1965calculation} to name a few. Another interesting property of Feynman integrals is that they can in general be written as multi-variable hypergeometric functions\cite{delaCruz:2019skx,Klausen:2019hrg,Ananthanarayan:2022ntm} and are thus multi-valued. Once such computations are carried out the analytic properties of the Feynman integrals can be readily obtained. Since such calculations are at times difficult to carry out, there may be value in revisiting methods of algebra, geometry and analysis, to obtain insights into the analytic structure of the integrals without evaluating them explicitly, as well as insights in general.  We believe that the present work is part of the effort to realize this goal.  To this extent, we believe that this is research at the Frontier.

In this work, we focus on understanding the analytic properties of the Feynman integrals without carrying the aforementioned computation. We will analyse the Feynman integrals at the integrand level so as to obtain their singularities. We know that the $n-$point functions are used to describe the scattering of $ n-$ particles. Also, to write down the dispersion relations\cite{barton1965introduction,todorov2014analytic,eden2002analytic} one needs to know the analytic structure of the amplitude in question to define the proper integration contour. There are various other works where analytic structure of Feynman integrals has been previously studied in various contexts. These include analysis of Landau equations and that of  physical region singularity\cite{Stapp:1967:FNP}, unitarity with two or three particle in the intermediate states \cite{gribov1962analytic}, study of singularities for physical examples such as pi-pi scattering \cite{kolkunov1961location}  and studies related to spectral representation and Mandelstam representation in perturbative field theory\cite{PhysRev.111.1187,petrina1964mandelstam}. Thus, the analytic properties of the Feynman integral are of much interest in high-energy physics. Once the analytic computation of Feynman integrals has been done these analytic properties can be easily obtained by analysing the result. However, we would like to focus on methods where such analysis can be carried out without evaluating the Feynman integrals explicitly and at the integrand level itself. Our focus is to extract the various singularities contained in them \cite{eden2002analytic}. Formally the analytic properties of Feynman integrals have been studied mainly using the following two tools:
\begin{itemize}
    \item Landau equations: It gives the condition for the occurrence of the singularities \cite{eden2002analytic,landau1959analytic,Zwicky:2016lka,Ananthanarayan:2018tog,Flieger:2022xyq,Stapp:1967:FNP}.
    \item Cutkosky rules: These rules \cite{cutkosky1960singularities} are used to compute the discontinuity of an amplitude.
\end{itemize}

In this work we attempt to study these singularities using tools developed in $1960s$ \cite{hwa1966homology}. However, Feynman integrals as they appear in literature are not suited to directly apply these tools \cite{hwa1966homology,fotiadi1965applications}. So as to apply these tools we would study these integrals in the compactified space. This is achieved by embedding both the integration cycle as well as the ambient space(the space to which the loop momenta belongs), associated with them into a compact space. Analysis of a simple unitary integral using this technique has already been presented in \cite{hwa1966homology}, though the treatment of Feynman integral has not been carried out citing their complicated nature. The other work where such treatment of Feynman integrals has been attempted is the work of Federbush\cite{federbush1965note}, though the analysis is restricted only to the singularities of the second kind for bubble and triangle integral at one-loop and the double-box integral at two-loop level. Our aim would be to extend the use of this analysis  and also describe the existing analysis in detail.
The method we would use can be briefly described as follows
\begin{itemize}
    \item Consider a one-loop Feynman integral in Euclidean 4-space of the following form\footnote{We stick to the notation of \cite{hwa1966homology}.}
    \begin{equation*}
       \int_{{\mathbb R}^4} d^4 k \frac{1}{\prod_i S_i(t, k)}
      \end{equation*}
where $S_{i}(t,k)$ are the Feynman propagators. We call the space to which loop momenta $k$ belongs as the ambient space. In our case, the ambient space is $\mathbb{C}^{4}$ which is not compact. The integration cycle $\mathbb{R}^{4} \subset \mathbb{C}^{4}$ is also not compact. 
    \item The Feynman integrals in the above form fail to be in the standard form \cite{fotiadi1965applications}, which is essential for further analysis using the present approach. The reason for this is that both the ambient space as well as the integration cycle are not compact.
    \item To bring the Feynman integral into standard form by compactifying both the integration cycle as well as the ambient space associated with it. The compactification procedure to be used has been described in \ref{appendix:compact}. This is similar to the compactification of complex plane $\mathbb{C}$ into the Riemann sphere. 
    \item After the compactification procedure we have transformed propagators $S_{i}(t,k) \rightarrow S_{i}(t,x)$. $S_{i}(t,x)$ are hyperplanes in the compactified space ${\CP}^5$. The singularities of the Feynman integrals are then obtained by analyzing the intersection of these planes in the non-general position \ref{appendix:ngpos}.
\end{itemize}

For ease of understanding the above procedure has been described for the one-loop integrals but it can be generalized to higher loops as well. As we have already mentioned, the method described above has been used to study the second-type singularities\cite{federbush1965note} in a few cases. With this motivation, we use the method to analyze other singularities viz threshold, pseudo-threshold and anomalous threshold using the method and show that the procedure allows us to study all the kinds of singularities in a single framework. We will also refine the analysis to further determine whether the singularities lie on the physical sheet or not for some tractable examples. For the simple case of the bubble and the triangle integral we will see that this analysis is very similar to the conditions on the Feynman parameters for the singularities to lie on the physical sheet \cite{Coleman:1965xm}.

The outline of the paper is as follows: In Section \ref{sec:landau} we review the Landau analysis using a simple example of a one-loop bubble integral. We also introduce some mathematical preliminaries such as non-general position and the compactification for both single as well as higher loops. These preliminaries are essential for the further development of the paper. In the subsequent Sections \ref{sec:bub}, \ref{sec:triangle}, \ref{sec:box}, \ref{sec:sunset} and \ref{sec:dbox} we analyse one-loop cases such as the bubble integral, the vertex integral, the box integral, as well as two-loop cases such as the two-loop sunset and the double-box integrals. We analyse various singularities associated with them extending the previous analysis. We also discuss the procedure to determine the singularities in the physical sheet. This is followed by a discussion of future work. To further fill the gaps in the calculations we have provided a \textsc{Mathematica} file \texttt{Calculation.nb} which can be found at: \href{https://github.com/TanayPathak-17/Singularities-of-Feynman-Integrals}{ GitHub}.
\section{Preliminaries}\label{sec:preliminary}
In this section, we discuss the preliminaries essential for the further analysis carried out in the subsequent sections.
\subsection{Landau analysis}\label{sec:landau}
We first briefly discuss how the singularities of the Feynman integrals are obtained using the Landau equations \cite{eden2002analytic,landau1959analytic,Zwicky:2016lka,Ananthanarayan:2018tog,Flieger:2022xyq}. 

A generic Feynman integral with L-loop of momenta $k_{i}(i=1,\cdots L)$, $N-$ propagators, external momenta $p_{i}$ can be written as follows
\begin{equation}
    I=\int D k \frac{1}{\prod_{i=1}^N\left(q_i^2-m_i^2\right)}, \quad D k=\prod_{i=1}^L d^4 k_i
\end{equation}
Using Feynman parameters we can write the above integral as follows
\begin{equation}\label{eq:feynpara}
I=\int D k \int_0^1 D \alpha \frac{1}{(F)^N}, \quad D \alpha=\prod_{i=1}^N d \alpha_i \delta\left(1-\sum_{i=1}^N \alpha_i\right)
\end{equation}
where 
\begin{equation}\label{eq:Fvalue}
  F =  \sum_{i=1}^{N}\alpha_{i}(q_{i}^{2}-m_{i}^{2})  
\end{equation}
and $\alpha_{i}$ are called the Feynman parameters.

The main idea for the analysis of singularities is that there are different classifications of singularities depending on how many of the $N$ propagators are on-shell, i.e. $q_{i}^2 = m_{i}^2$. This idea is more concretely given by Landau equations which tell that the singularities occur when 
 \begin{enumerate}
     \item $q_{i}^2 = m_{i}^2$\label{lequation1}
     \item There exists $\alpha_i$, not all $0$, such that $\sum_{i \in \operatorname{loop}(l)} \alpha_i\left(q_i\right)^\mu=0 \text { for loop } l=1 \ldots L$\label{lequation2} 
 \end{enumerate}

 The Landau equation \ref{lequation1}, $q_{i}^2 = m_{i}^2$ implies that the corresponding propagator is on-shell. This in turn assures that $F=0$ in Eq. \eqref{eq:feynpara}, by demanding that each term in the summand \eqref{eq:Fvalue} is zero. The Landau equation \eqref{lequation2}, can be interpreted in a geometric manner. It tells us that the corresponding singularity surfaces are parallel to each other and the hypercontour cannot be deformed away from the approaching singularity surfaces \cite{Zwicky:2016lka}. Furthermore, if $\alpha_{i}=0$ for any $i$, it means that the corresponding propagator does not contribute to the singularity. The singularity corresponding to $\alpha_{i} \neq 0$, for all $i$, is called the leading singularity. All others are called sub-leading singularities.  
 
 As an example, we consider the simple case of a one-loop two-point function. The bubble Feynman integral is given by 
 \begin{equation*}
    I_{B} = \int \frac{d^{4}k}{(k^{2}-m_{1}^{2})((k-p)^{2}-m_{2}^{2})}
\end{equation*}
hence we have $q_{1}= k$ and $q_{2}=k-p$. The first Landau equation gives $q_{1}^2 = m_{1}^{2}$ and $q_{2}^{2} = m_{2}^{2}$. The second Landau equation can be cast into the form $\det(Q) =0$, where
\begin{align}
    \det (Q)=\operatorname{det}\left(\begin{array}{cc}
m_1^2 & q_1 \cdot q_2 \\
q_1 \cdot q_2 & m_2^2
\end{array}\right)=0 \nonumber \\
 q_{1}\cdot q_{2} = \pm m_{1}m_{2}
\end{align}
Reinserting  into $p=q_{1} - q_{2}$, yields following two singularities\\
\begin{center}
    $p^{2}_{(+)} = (m_{1}+m_{2})^2$ and $p^{2}_{(-)} = (m_{1}-m_{2})^{2}$\\
\end{center} 
To determine in which sheets the above singularities lie, further analysis has to be done and the Landau equations have to be refined. The analysis reveals that the singularity $p_{(+)}^{2}$ lies on the physical sheet while the singularity $p_{(-)}^{2}$ does not. These singularities are called threshold and pseudo-threshold singularities respectively. They are also the leading singularities for the present case. 

Following \cite{Zwicky:2016lka}, we can also look at the geometrical interpretation of the above singularities. The two Landau equations give
\begin{center}
    $k^{2} = m_{1}^{2}$, \quad $(k-p)^{2}=m_{2}^{2}$
\end{center}
These two equations defines two hyperboloids with their centres displaced by $p$ (see Fig.9,\cite{ Zwicky:2016lka}). Then for any light-like $p$, these two hyperboloids meets at infinity, thus giving rise to second type singularities. Another interpretation of the origin of these singularities is that they arise due to pinching at infinity \cite{eden2002analytic}.

The second-type singularities is determined by looking at the vanishing of the Gram-determinant 
\begin{equation}
    \det \, p_{i} \cdot p_{j} = 0 
\end{equation}
For the case of one-loop bubble integral the above condition gives $p^{2}=0$. These singularities are independent of the masses and do not lie in the physical sheet. Using the homological techniques we will see that all three types of singularities of the bubble integral can be incorporated within a single framework and no special analysis is required to obtain the second type of singularity.

\subsection{Compactification}\label{appendix:compact}
The detailed compactification procedure is described in \cite{hwa1966homology}. A Feynman integral in 4-space has the form \footnote{We assume that in all the Feynman integrals, we consider the parameters and integration variable that appear are ``dimensionless" quantities which are divided by some fundamental `mass' relevant to a particular theory under consideration.}
\begin{equation}\label{compfeyn}
\int_{{\mathbb R}^4} d^4 k \frac{1}{\prod_i S_i(t, k)}
\end{equation}

There are two problems here. First, the ambient space  $\mathbb{C}^{4}$ is not compact and second, the domain of integration   $\mathbb{R}^{4}$ is also not compact.  For further calculations, it is useful for both of them to be compact. We follow \cite{hwa1966homology}

We compactify the ambient space ${\mathbb C}^4$ by embedding it in to $\CP^5$ using the map 
\begin{align}\label{eq:compact}
    {\mathbb C}^4 \longrightarrow \CP^5 \nonumber\\
    k \longrightarrow \mathbf{x}:=(2k,1-k^2,1+k^2)
\end{align}
That is $x_i=2k_i, 1\leq i \leq 4, x_5=1-(k_1^2+k_2^2+k_3^2+k_4^2)$ and $x_6=1+(k_1^2+k_2^2+k_3^2+k_4^2)$. We could instead embed $\C^4$ into $\CP^4$ but certain computations become clearer this way.

Under this map, the domain of integration $\R^4$ is taken to the set $(2k,1-k^2,1+k^2)$ in $\CP^5$. Since $k\in \R^4$, $1+k^2$ is always non-zero. Hence we can divide by $(1+k^2)$ and consider it as a subset of $\R^5 \subset \C^5$, where $\C^5$ is the subset of $\CP^5$ given by $x_6=1$. So the closure of the image of $\R^4$ is the closure of the image of the set
$\left(\frac{2k}{1+k^2},\frac{1-k^2}{1+k^2}\right) \in \R^5.$
This is the unit sphere and is compact. The above compactification procedure is essentially the inverse stereographic projection \cite{arfken2011mathematical} and further homogenization of the resulting coordinates \cite{silverman1992rational}.

Recall that the denominators of the Feynman integral \eqref{compfeyn} are of the form $S_{i}(t,k)= (a_{i}(t)+k)^{2} - m_{i}^{2}$. Under the mapping above, these are  taken to 
\begin{equation}\label{denocom}
    S_{i}(t,k) \longrightarrow S_{i}(t,\mathbf{x}) = \frac{\mathbf{x} \cdot \mathbf{A}_{i}}{x_{5}+x_{6}}
\end{equation}
where $\mathbf{A}_{i}= (2a_{i}(t),a_{i}^{2}-m_{i}^{2}-1,a_{i}^{2}-m_{i}^{2}+1)$. The dot product is defined as 
\begin{align*}
\mathbf{x} \cdot \mathbf{y} = x_{1}y_{1} +x_{2}y_{2}+x_{3}y_{3}+x_{4}y_{4}+x_{5}y_{5}+x_{6}y_{6} 
\end{align*} Similarly, $d^{4}k$ is taken to  
\begin{equation}\label{meascom}
d^4 k \longrightarrow\left(\frac{1}{x_5+x_6}\right)^4 d x_1 \wedge d x_2 \wedge d x_3 \wedge d x_4
\end{equation}
We remark that in Eq.\ref{denocom} the compactified propagator is homogeneous, i.e. it is invariant under any transformation of the form $\mathbf{x} \rightarrow \lambda \mathbf{x} $, where $\lambda$ is some scalar.  From Eq. \eqref{denocom} and \eqref{meascom} we can further see that if there are less than four propagators in the Feynman integral we have an effective denominator in the compactified integral. This denominator corresponds to surface $S_{0} \equiv x_{5}+x_{6}=0$ and is singular. The singularities arising due to the intersection of $S_{0}$ and $S_{i}$ in non-general position give rise to second-type singularities.

In the case of two or higher loop integrals, there is no general formula for a generic denominator as \eqref{denocom}. In these cases, one has to recursively apply \eqref{denocom} for each of the loop momenta $k_{i}$ thus compactifying each of the $k_{i}$ into a copy of $\CP^5$.

As an example consider the two-loop Sunset integral \eqref{sunsetint}. It has the following three propagators
\begin{center}
 $k_1^2-m_1^2, \quad k_2^2-m_2^2$ and $(k_1 +k_2-p)^2 -m_3^2$ \\   
\end{center}
The first two propagators are easily dealt with using Eq.\eqref{denocom} and we get 
\begin{align*}
S_{1}&= \frac{x_{5}(-m_{1}^{2}-1)+x_{6}(-m_{1}^{2}+1)}{x_{5}+x_{6}} \\
S_{2}&= \frac{y_{5}(-m_{2}^{2}-1)+y_{6}(-m_{2}^{2}+1)}{y_{5}+y_{6}}
\end{align*}
For the third propagator,  we have to use  Eq. \eqref{denocom} twice, once for each $k_{i}$. We can write the third propagator in the following suggestive form
\begin{equation*}
    S_{3} =  (k_{1}^{2}-m_{3}^{2})+ (k_{2}-p)^{2} + \frac{2 (2k_{1})\cdot(2k_{1}-2 p)}{4}
\end{equation*}
Using Eq.\eqref{denocom} for each of the pieces and simplifying,  we get
\begin{equation*}
    S_{3} = \frac{(y_{5}+y_{6})(x_{5}(-m_{3}^{2}-1)+x_{6}(-m_{3}^{2}+1))+(-2 y \cdot p + y_{5}(p^{2}-1)+y_{6}(p^{2}+1))(x_{5}+x_{6}) }{(x_{5}+x_{6})(y_{5}+y_{6})}
\end{equation*}
which is the required compactification. We emphasise the fact that to keep the $S_{3}$ invariant under the transformation $\mathbf{x} \rightarrow \lambda_{1} \mathbf{x}, \mathbf{y} \rightarrow \lambda_{2} \mathbf{y} $, the factor of $(x_{5}+x_{6})(y_{5}+y_{6})$ in the denominator is important.

\subsection{Non-general position}\label{appendix:ngpos}
\begin{figure}[t]
		\centering
		\begin{subfigure}[b]{0.48\textwidth}
		\centering
		\includegraphics[width=\textwidth]{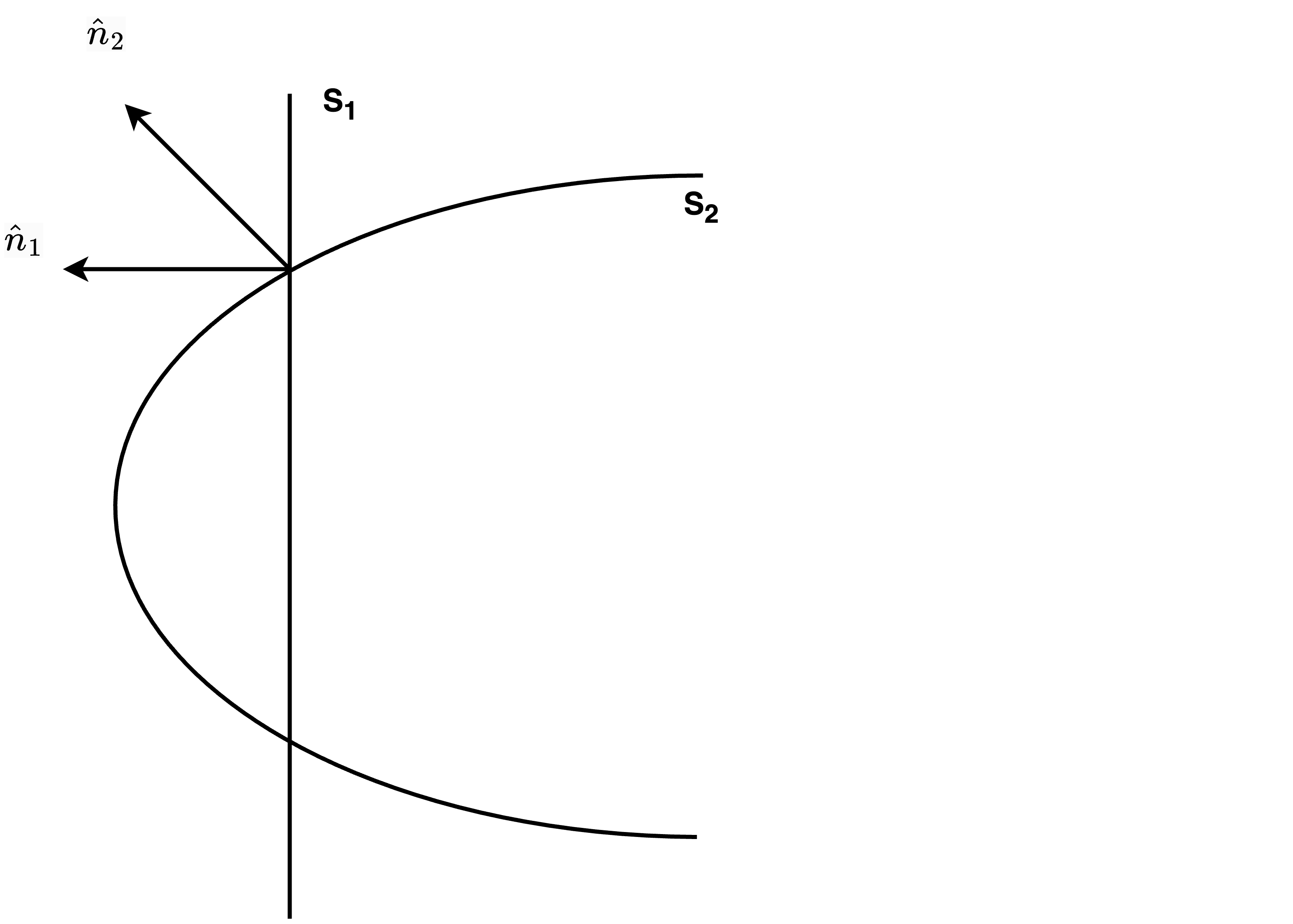}
			\caption{}
			\label{fig:generalpos}
		\end{subfigure}
		\hfill
		\begin{subfigure}[b]{0.50\textwidth}
		\centering
		\includegraphics[width=\textwidth]{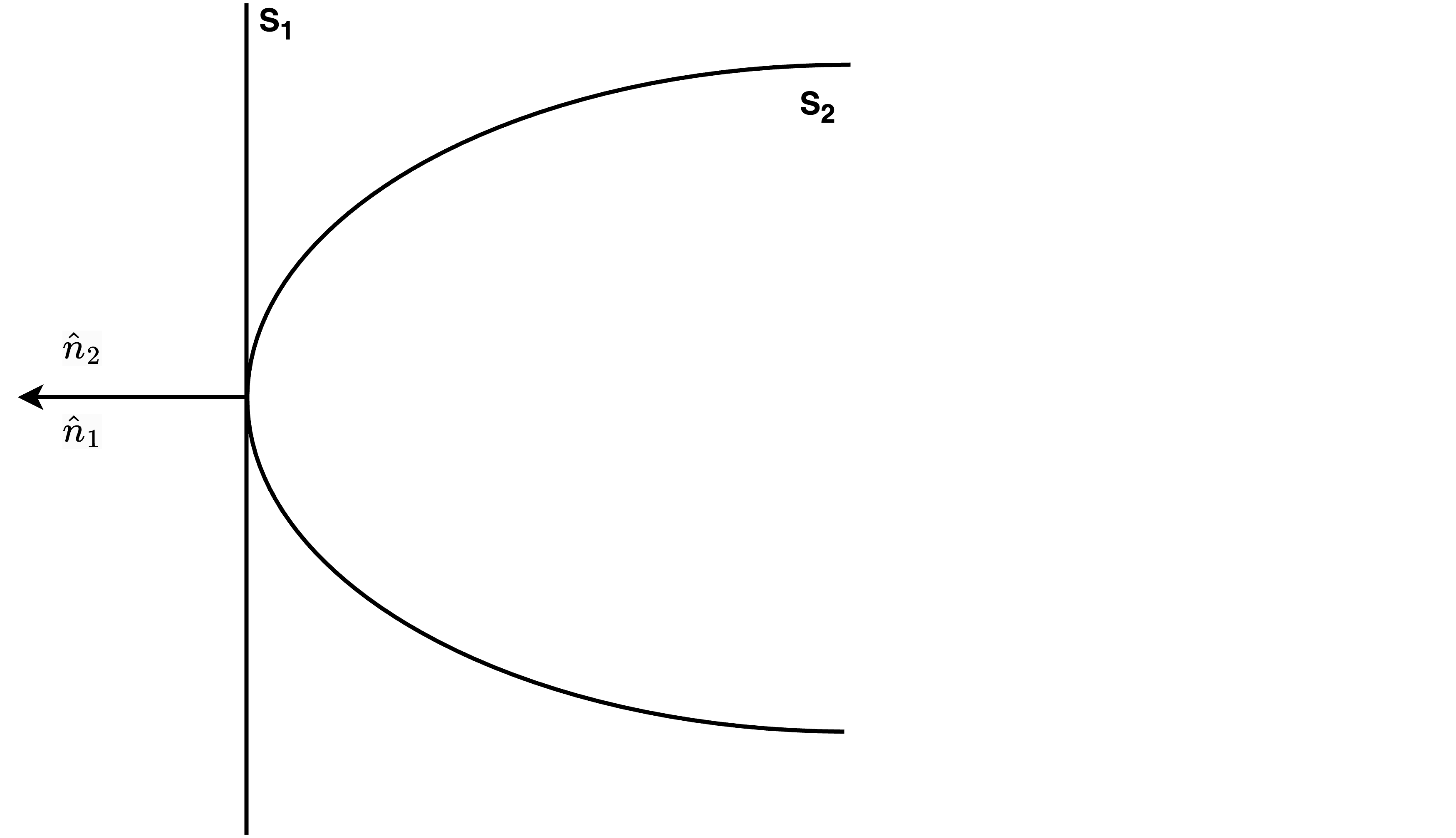}
			\caption{}
			\label{fig:nongeneralpos}
		\end{subfigure}
		\caption{(a) Two surfaces meeting in general position (b) Two surfaces meeting in non-general position. We see that for the case of non-general position, at the point of intersection, the normal to both surfaces are parallel to each other. }
		\label{kcplots1}
\end{figure}
In this subsection, we briefly outline the concept of non-general position as given in \cite{hwa1966homology}. Surfaces $S_{i}$ in non-general position intersect at a  simple pinch giving rise to the singularities we are interested in.

A pinch is only possible when the surfaces meet at non-general position, which implies that the following conditions are satisfied \footnote{We introduce  parameters $\alpha_{i}$ sticking to the notation in \cite{hwa1966homology}, they are not same as the Feynman parameters.} 
\begin{enumerate}
    \item  $S_{i} = 0 , i= 1, \cdots m$ \label{ngcondition1}
    \item There exists $\alpha_i$, not all $0$,  such that  $\sum_{i=1}^m \alpha_i \frac{\partial S_i}{\partial x_k}=0$, $k= 1, \cdots , l$ \label{ngcondition2}
\end{enumerate}
In the case of Feynman integrals $S_{i}$ will be the compactified propagators obtained via compactification described in \ref{appendix:compact} and the conditions \ref{ngcondition1} and \ref{ngcondition2} are valid for any $L-$loop Feynman integral. 
We further notice that the above conditions bear resemblance to Landau equations described in sub-section \ref{sec:landau}. Though it is to be mentioned that the above conditions are more general consideration and valid for any family of hyper-surfaces $S_{i}$- the Landau equations can be thought of as a special case of the same when applied to Feynman integrals.

Geometrically, condition \ref{ngcondition1} restricts $x$ to $S_{1} \cap S_{2} \cap \cdots \cap S_{m}$ and condition \ref{ngcondition2} implies that the normal vectors to $S_{i}$ at $x$ are linearly dependent.
For the case when $m=2$, Condition 2 implies that the two normals are parallel, see Fig. \ref{fig:generalpos} and \ref{fig:nongeneralpos}. Similarly, when $m=3$ this implies that the three normals are co-planar.

\section{One loop Bubble integral}\label{sec:bub}

From this section onwards, we analyse various one and two-loop integrals. We have also considered a toy example to demonstrate the method in a lower-dimensional integral in appendix \ref{appendix:toy}. Due to the tedious calculation involved at times, we have provided a \textsc{Mathematica} file \texttt{Calculation.nb} to fill the gaps for the reader. The file can be found at \href{https://github.com/TanayPathak-17/Singularities-of-Feynman-Integrals}{GitHub}.
\begin{figure}[H]
\centering\includegraphics[width=.35\textwidth]{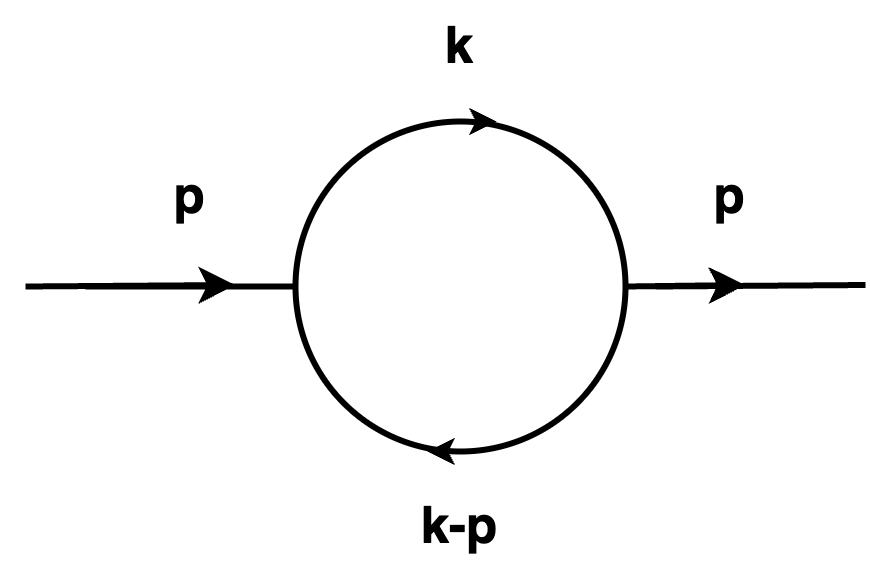}
\caption{Bubble diagram}\label{figbubble}
\end{figure}
Let us consider the one-loop Bubble Feynman integral corresponding to the diagram in Fig.\ref{figbubble} 
\begin{equation}\label{intbub}
    I_{2} = \int \frac{d^{4}k}{(k^{2}-m_{1}^{2})((k-p)^{2}-m_{2}^{2})}
\end{equation}
To compactify the propagators, we make the following transformation as given in Eq. 
 \eqref{eq:compact}
\begin{equation}
    x_{\alpha}= 2 k_{\alpha},\quad x_5= 1-k^{2}, \quad x_6= 1+k^{2}
\end{equation}
Using Eq.\eqref{denocom} we get the following compactified propagators
\begin{align}\label{propb}
    S_1 &= \frac{x_{5}(-m_{1}^{2}-1)+ x_{6}(-m_{1}^{2}+1)}{x_{5}+x_6}\nonumber \\ 
    S_2&= \frac{-2 p\cdot x +x_{5}(p^{2}-m_{2}^{2}-1)+x_{6}(p^{2}-m_{2}^{2}+1)}{x_{5}+x_6}\\ \nonumber
\end{align}
and the new ambient space is $W$ given by: 
\begin{equation*}
    W= \{(x_{1},\cdots,x_{6})|\sum_{i}^{5}x_{i}^{2}-x_{6}^{2}=0 \}\subset \mathbb{CP}^5
\end{equation*}. 
We also have an effective denominator $S_{3}= x_{5}+x_{6}$ which can be identified as the plane at infinity. It will be later shown that the intersection of the plane at infinity with the compactified propagators($S_{1}$ and $S_{2}$ in the present case) give rise to singularities of the second kind. 

In the compactified space the integral is 
\begin{equation}
I_{2}= \int \frac{dx_{1}dx_{2}dx_{3}dx_{4}}{(x_{5}+x_{6})^{2} S_{1}S_{2}}
\end{equation}
where $S_1$ and $S_2$ are given by Eq.\eqref{propb}.

The singularities corresponding to the integral above are given when the denominators $S_{1}$,$S_{2}$ and $S_{3}$ meet at non-general position in $W$, see subsection \ref{appendix:ngpos}. 
We first analyse the case when $S_1$ and $S_2$ are in a non-general position in $W$. Using the condition \ref{ngcondition1} for surfaces meeting in a non-general position we get 
\begin{align}\label{eqalphabub1}
-x_{5}+x_{6} &= (x_{6}+x_{5})m_{1}^{2}\nonumber \\
    (x_{5}+x_{6})(p^{2} -m_{2}^{2})+(x_{6}-x_{5}) &= 2 p\cdot x \nonumber \\
   \sum_{i}^{5}x_{i}^{2}-x_{6}^{2}&=0
\end{align}
 Using the condition \ref{ngcondition2} we get 
\begin{align}\label{eqalphabub2}
\alpha_{2}(-2p) +\alpha_{3}(2 x)&=0 \nonumber \\
\alpha_{1}(-m_{1}^{2}-1)+\alpha_{2}(p^{2}-m_{2}^{2}-1)+\alpha_{3}(2x_{5})&=0 \nonumber \\
\alpha_{1}(-m_{1}^{2}+1)+\alpha_{2}(p^{2}-m_{2}^{2}+1)+\alpha_{3}(-2x_{6})&=0 
\end{align}
Using Eq.\eqref{eqalphabub1}, we get 
\begin{equation}
    2 p\cdot x= (x_{5}+x_{6})(p^{2}-m_{2}^{2}+m_{1}^{2})
\end{equation}
We perform the dot product of $p$ in the first relation of Eq.\eqref{eqalphabub2}, . This converts the equation into an equation with scalar coefficients whose value we know. Performing this we get 
\begin{equation}
    \alpha_{2}(-2p^{2}) +\alpha_{3}(2 p \cdot x) =0
\end{equation}
substituting the value of $2 p \cdot x$ we get 
\begin{equation}
    \alpha_{2}(-2p^{2}) +\alpha_{3}((x_{5}+x_{6})(p^{2}-m_{2}^{2}+m_{1}^{2})) =0
\end{equation}
With this Eq.\eqref{eqalphabub2} becomes
\begin{align}
   \alpha_{2}(-2p^{2}) +\alpha_{3}((x_{5}+x_{6})(p^{2}-m_{2}^{2}+m_{1}^{2})) &=0 \nonumber \\
\alpha_{1}(-m_{1}^{2}-1)+\alpha_{2}(p^{2}-m_{2}^{2}-1)+\alpha_{3}(2x_{5})&=0 \nonumber \\
\alpha_{1}(-m_{1}^{2}+1)+\alpha_{2}(p^{2}-m_{2}^{1}+1)+\alpha_{3}(-2x_{6})&=0 
\end{align}
We want the nontrivial solution for $\alpha_{i}$ in equations the above equation, which is possible when the matrix of the coefficients of $\alpha_{i}$ s has a vanishing determinant. That is, 
\begin{align}
     \left| \begin{array}{ccc}
 0&-2 p^2 & ((x_{5}+x_{6})(p^{2}-m_{2}^{2}+m_{1}^{2})) \\
 -m_{1}^{2}-1 \quad & p^{2}-m_{2}^{2}-1 & 2x_{5} \\
 -m_{1}^{2}+1 & p^{2}-m_{2}^{1}+1 & -2x_{6} \end{array}
\right|=0
\end{align}

Evaluating the above determinant and solving for the invariant $p^{2}$.
\begin{equation}\label{bubsethreholpseudo}
    p^{2}= (m_{1}-m_{2})^{2},(m_{1}+m_{2})^{2}
    \end{equation}
These are precisely the conditions for the threshold and pseudo-threshold singularity.

Next, we consider the case of the second type singularity for the Bubble integral, which has been analysed in \cite{federbush1965note} as well. The second type of singularity occurs when $S_1$, $S_2$ and $S_3$ are in non-general position and $S_{1}$ and $S_{2}$ meet in general position. Using the first condition for surfaces meeting in a non-general position we get 
\begin{align}
    x_{6}+x_{5}&=0 \nonumber\\
    -x_{5}+x_{6} &= (x_{6}+x_{5})m_{1}^{2}\nonumber \\
    (x_{5}+x_{6})(p^{2} -m_{2}^{2})+(x_{6}-x_{5}) &= 2 p\cdot x \nonumber \\
   \sum_{i}^{5}x_{i}^{2}-x_{6}^{2}&=0
\end{align}
Using the second condition we get
\begin{align}
    \alpha_{1}(x) + \alpha_{2}(-p)&=0
\end{align}
Performing dot product with $p$ in the above we obtain
\begin{align}
    \alpha_{1}(p \cdot x) + \alpha_{2}(-p^{2})&=0
\end{align}
For $\alpha_{i}$ to have a non-trivial solution, we need the following
\begin{equation}\label{bubsecond}
    p^{2}=0
\end{equation}
 which gives the second-type singularity. We see that in this picture, the second type singularity occurs due to the intersection of the planes at infinity ($S_{3}$ in the present case) with the other planes ($S_{1}$and $S_{2}$ in the present case). In the case of the Bubble integral, we summarize all the results in Table \ref{tablebubsing}.
\begin{table}[]
\centering
\begin{tabular}{|l|l|l|}
\hline
Type & Singularity & Equation \\ \hline
    Threshold  &   $p^{2}=(m_{1}+m_{2})^{2}$          &     Eq.\eqref{bubsethreholpseudo}      \\ \hline
   Pseudo-threshold  &  $p^{2}=(m_{1}-m_{2})^{2}$& Eq.\eqref{bubsethreholpseudo}\\ \hline
    Second-type & $p^{2}=0$ &   Eq.\eqref{bubsecond}\\ \hline
\end{tabular}
\caption{Singularities of Bubble integral Eq.\eqref{intbub}}\label{tablebubsing}
\end{table}

\begin{center}
    \textbf{Properties of $\alpha_{i}$ at the singularity}
\end{center}

We now discuss some features of the parameters $\alpha_{i}$ at the singularity. Using Eq.\eqref{eqalphabub1} and \eqref{eqalphabub2} we  get the following values of $\alpha$
\begin{equation}
    \alpha_1= -\frac{2 \alpha_3 x_3 \left(m_1^2+m_2^2-p^2\right)}{\left(m_1^2+1\right) \left(m_1^2-m_2^2+p^2\right)}, \quad \alpha_2= \frac{4 \alpha_3 m_1^2 x_3}{\left(m_1^2+1\right) \left(m_1^2-m_2^2+p^2\right)}
\end{equation} Witout loss of generality, we may assume $m_1>m_2$ and $\alpha_3=x_3=1$. We consider the following three cases:
\begin{enumerate}
 \item Threshold : For the threshold singularity $p^{2}=(m_{1}+m_{2})^2$. We then  get the following value of $\alpha_1$ and $\alpha_2$
 \begin{equation}
    \alpha_{1}= \frac{2 m_2}{\left(m_1^2+1\right) \left(m_1+m_2\right)},  \quad \alpha_{2}=\frac{2 m_1}{\left(m_1^2+1\right) \left(m_1+m_2\right)}
 \end{equation}
 We see that both are positive.

  \item Pseudo-threshold : For the pseudo-threshold singularity, $p^{2}=(m_1 -m_2)^{2}$.  We get
  \begin{equation}
    \alpha_{1}= -\frac{2 m_2}{\left(m_1^2+1\right) \left(m_1-m_2\right)},\quad \alpha_{2}=\frac{2 m_1}{\left(m_1^2+1\right) \left(m_1-m_2\right)}
  \end{equation}
Notice that $\alpha_{1}$ is negative and $\alpha_{2}$ is positive.

\item Second type singularity: For second-type singularity, $p^2=0$. We get
\begin{equation}
   \alpha_{1}= -\frac{2 \left(m_1^2+m_2^2\right)}{\left(m_1^2+1\right) \left(m_1^2-m_2^2\right)},\quad \alpha_{2}=\frac{4 m_1^2}{\left(m_1^2+1\right) \left(m_1^2-m_2^2\right)}
\end{equation}
Notice that again $\alpha_{1}$ is negative and $\alpha_{2}$ is positive.
\end{enumerate}

From the conventional analysis, it is known that the threshold singularities lie on the physical sheet and the pseudo-threshold and the second-type singularities do not lie on the physical sheet. We observe that it is only in the case of the threshold singularity that the values of $\alpha_{1}$ and $\alpha_{2}$ are both positive. Hence if the sign of $\alpha_{i}$ is positive at a singularity then the singularity is in the physical sheet. We would like to emphasise that this feature is similar to that of Feynman parameters, which have to be positive for the singularities to lie on the physical sheet. In \cite{Coleman:1965xm} it has been shown using physical arguments that the Feynman parameters have to be positive for singularities to lie on the physical sheet. The similar feature of $\alpha_{i}$s, thus hints towards the connection between the two.

\section{Triangle integral}\label{sec:triangle}
\begin{figure}[H]
\centering\includegraphics[width=.35\textwidth]{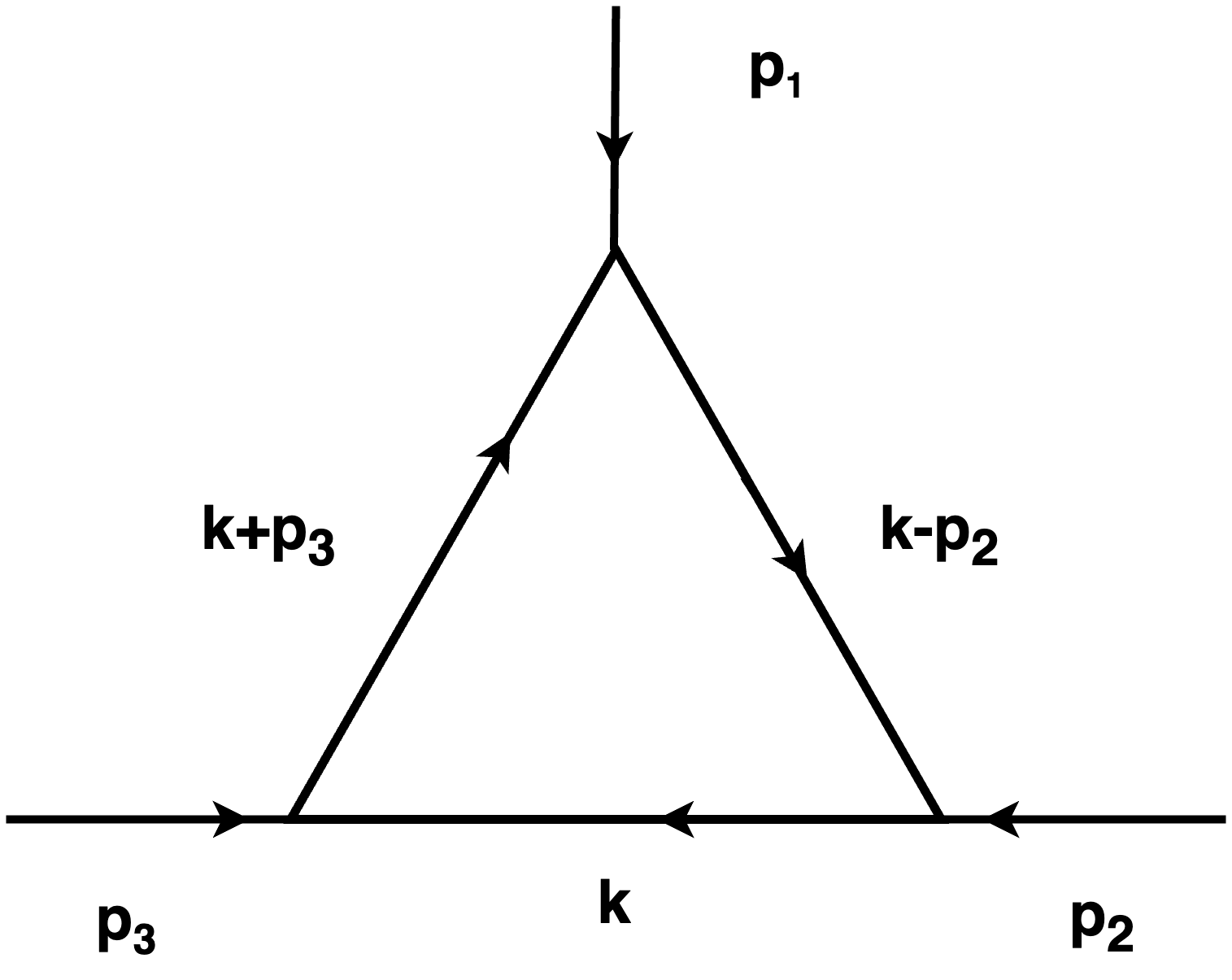}
\caption{Triangle diagram}\label{figtriangle}
\end{figure}
We now consider the Triangle Integral corresponding to the triangle diagram in Fig.\ref{figtriangle}
\begin{equation}
I_{3}= \int \frac{d^{4}k}{(k^{2}-m_{1}^{2})((k+p_{3})^{2}-m_{2}^{2})((k-p_2)^{2} -m_{3}^{2})}
\end{equation}
The three propagators in the triangle diagram are given as follows:
$$k^{2}-m_{1}^{2}, \quad (k+p_{3})^{2}-m_{2}^{2},\quad (k-p_2)^{2} -m_{3}^{2} $$
Compactifying the propagators as in the previous case, we get the following 
\begin{align}
    S_1 &= \frac{x_{5}(-m_{1}^{2}-1) +x_{6}(-m_{1}^{2}+1)}{x_{5}+x_{6}} \nonumber \\
    S_{2} &= \frac{2 p_{3}\cdot x +x_{5}(p_{3}^{2}-m_{2}^{2}-1)+x_{6}(p_{3}^{2}-m_{2}^{2}+1)}{x_{5}+x_{6}} \nonumber \\
    S_{3} &= \frac{2 (p_3 +p_1) +x_{5}((p_3 +p_1)^{2} -m_{3}^{2}-1)+x_{6}((p_3 +p_1)^{2} -m_{3}^{2}+1)}{x_{5}+x_{6}} \nonumber \\
\end{align} 
and the new ambient space $W$ is given by
\begin{equation*}
    W= \{(x_{1},\cdots,x_{6})|\sum_{i}^{5}x_{i}^{2}-x_{6}^{2}=0 \}\subset \mathbb{CP}^5
\end{equation*}
We also get an effective denominator as 
\begin{equation}
     S_4 = x_{5} + x_{6} 
\end{equation}

The analysis is similar to the previous section. The leading singularity of the triangle integral is called the pseudo-threshold singularity. This occurs when $S_1, S_2$ and $S_3$ meet at non-general position in $W$. We get the following condition for the anomalous threshold singularity 
\begin{align}\label{triangathresh}
    \left|
\begin{array}{cccc}
 0 & 2 p_3^2 & -p_1^2+p_2^2+p_3^2 & -\frac{\left(1-m_1^2\right) \left(-m_2^2+p_3^2-1\right)}{m_1^2+1}+m_2^2-p_3^2-1 \\
 0 & -p_1^2+p_2^2+p_3^2 & 2 p_2^2 & -\frac{\left(1-m_1^2\right) \left(-m_3^2+p_2^2-1\right)}{m_1^2+1}+m_3^2-p_2^2-1 \\
 -m_1^2-1 & -m_2^2+p_3^2-1 & -m_3^2+p_2^2-1 & \frac{2 \left(1-m_1^2\right)}{m_1^2+1} \\
 1-m_1^2 & -m_2^2+p_3^2+1 & -m_3^2+p_2^2+1 & -2 \\
\end{array}
\right|=0
\end{align}
This matches the result given in \cite{eden2002analytic}, where it was obtained using the Feynman parameterized form of the Triangle integral.

We can further simplify this result to compare it with other literature results \cite{Zwicky:2016lka}. We take the following values $p_{2}=p, p_3=p, m_2=m$ and $ m_3=m$. With these special values, we get
\begin{equation}\label{athreholdsim}
   p_{1}^{2}= 4 m^2-\frac{(-m^2-m_1^2+p^2)^2}{m_1^2}
\end{equation}
The above singularity lies below the two-particle threshold $4m^2$ and is called the pseudo-threshold singularity.

Now let us analyse second-type singularities for the Triangle diagram. There are two cases when these singularities can occur.
\begin{enumerate}
    \item The first case arises when the sets $\{S_{1}, S_{2},S_{4}\}$,$\{S_{1}, S_{3},S_{4}\}$ and $\{S_{2}, S_{3},S_{4}\}$ are in non-general position in $W$. This is similar to the case of one loop bubble integral and hence we simply get the singularity 
    \begin{align}\label{trainsecond1}
       p_{3}^2=0,\quad p_{2}^2=0,\quad p_{1}^2=0 
    \end{align} 
    \item The second case arises when $S_{1}, S_{2}, S_{3}$ and $S_{4}$ are in non-general position in $W$. 
Analyzing this case in a similar manner as before we obtain the following singularity:
\begin{align}\label{trainsecond2}
    p_{1}^{2} p_{2}^{2} = (p_{1} \cdot p_{2})^{2}
\end{align}
The second type of singularity for the triangle has also been obtained in \cite{federbush1965note}.
\end{enumerate}
For the triangle integral, we thus have the following singularities 
\begin{table}[H]
\centering
\begin{tabular}{|l|l|}
\hline
Type & Equation \\ \hline
   Anomalous threshold &   Eq. \eqref{triangathresh} and \eqref{athreholdsim}       \\ \hline
    Second type &  Eq.\eqref{trainsecond1} and \eqref{trainsecond2} \\ \hline
\end{tabular}
\end{table}

\begin{center}
    \textbf{Properties of $\alpha_{i}$ at the singularity}
\end{center}
We will now study the properties of the parameters $\alpha_{i}$. For the present case, we will consider the leading singularity because of its interesting feature. We consider the simpler special case of Eq.\eqref{athreholdsim}. For this case, we get the following values of $\alpha_{i}$
    \begin{align}
        \alpha _1&= \frac{2 \left(p^{2}-m^2-m_1^2\right)}{\left(m_1^2+1\right) \left(-m^2+m_1^2+p^2\right)}, \quad \alpha _2= \frac{2 m_1^2}{\left(m_1^2+1\right) \left(-m^2+m_1^2+p^2\right)}, \nonumber \\
        \alpha _3&= \frac{2 m_1^2}{\left(m_1^2+1\right) \left(-m^2+m_1^2+p^2\right)}
    \end{align}
We immediately observe that when $p^{2}>m^{2}+m_{1}^{2}$, $\alpha_{i}$ are all positive. When we have $p^{2}< m^{2}+m_{1}^{2}$, $\alpha_{1}$ has a negative sign and the other two have a positive sign. The former case corresponds to the singularity in the physical sheet and later corresponds to the singularity lying in the unphysical sheet\cite{Zwicky:2016lka}.

\section{Box Integral}\label{sec:box}
\begin{figure}[H]
\centering\includegraphics[width=.35\textwidth]{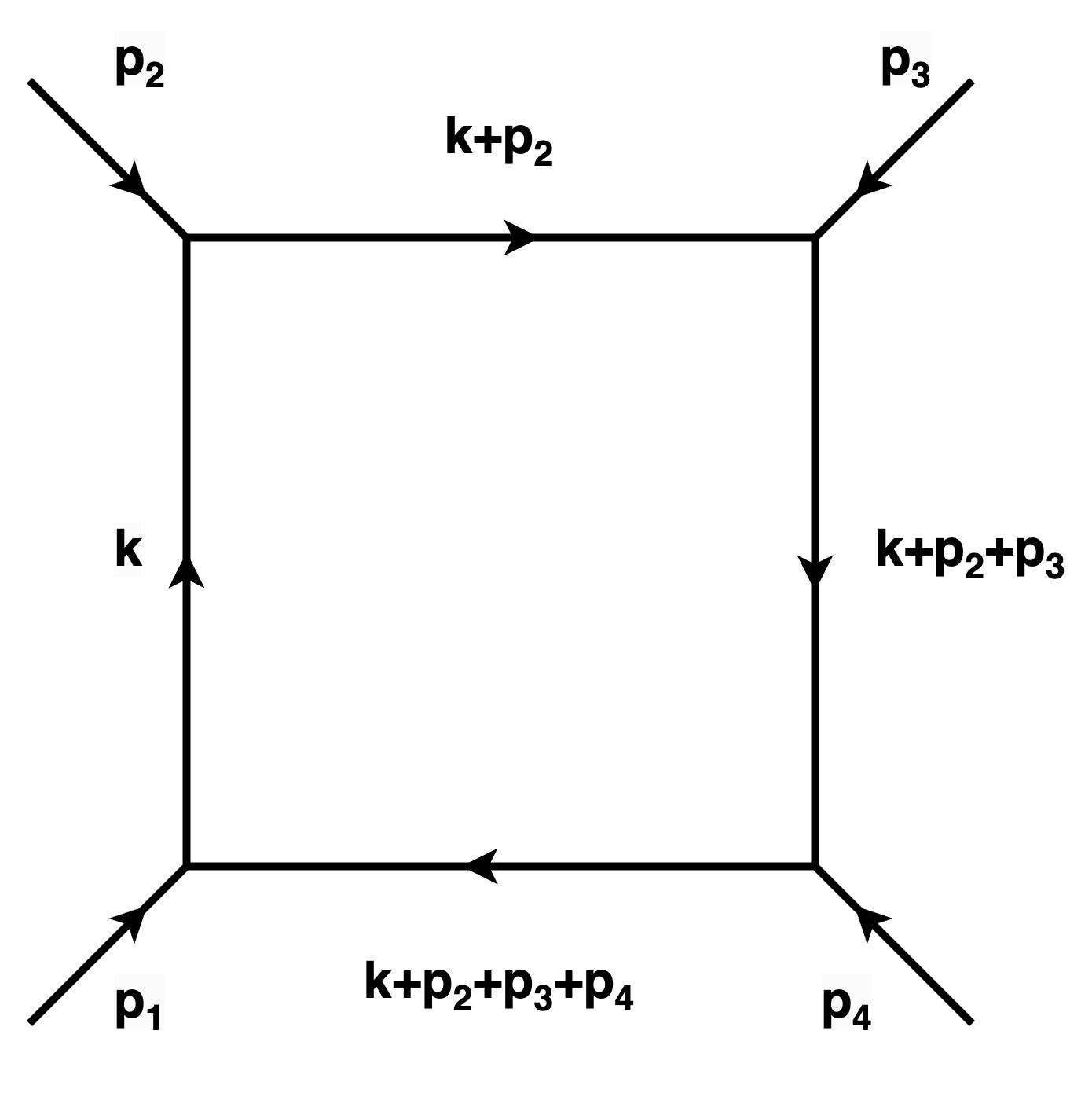}
\caption{Box diagram}\label{figbox}
\end{figure}
We now consider the Box Integral corresponding to the box diagram in Fig.\ref{figbox}
\begin{equation}
I_{4}= \int \frac{d^{4}k}{(k^{2}-m_{1}^{2})((k+p_{2})^{2}-m_{2}^{2})((k+ p_2 +p_3)^{2} -m_{3}^{2})((k+ p_2 +p_3+p_4)^{2} -m_{3}^{2})}
\end{equation}
The compactified propagators are as follows
\begin{align}
    S_{1}(t,x) &= \frac{(-m_{1}^{2}-1)x_{5}+(1-m_{1}^{2})x_{6}}{x_{5}+x_{6}} \nonumber \\
    S_{2}(t,x) &= \frac{2p_{2}\cdot x +(p_{2}^{2}-m_{2}^{2}-1)x_{5}+(p_{2}^{2}+1-m_{1}^{2})x_{6}}{x_{5}+x_{6}} \nonumber \\
    S_{3}(t,x) &= \frac{2(p_{2}+p_{3})\cdot x ((p_{2}+p_{3})^{2}-m_{1}^{2}-1)x_{5}+((p_{2}+p_{3})^{2}+1-m_{1}^{2})x_{6}}{x_{5}+x_{6}} \nonumber \\
    S_{4}(t,x) &= \frac{-2(p_{1} \cdot x)+(p_{1}^{2}-m_{1}^{2}-1)x_{5}+(p_{1}^{2}+1-m_{1}^{2})x_{6}}{x_{5}+x_{6}} 
\end{align}
where in $S_{4}$ we have used the momentum conservation $p_{1}+p_{2}+p_{3}+p_{4}=0$ condition. The new ambient space is $W$ given by: 
\begin{equation*}
    W= \{(x_{1},\cdots,x_{6})|\sum_{i}^{5}x_{i}^{2}-x_{6}^{2}=0 \}\subset \mathbb{CP}^5
\end{equation*}
In this case, we will not have any effective denominator as the number of propagators is four.

The singularities can arise in the following cases
\begin{enumerate}
    \item When two $S_{i}$ are in  non-general position in $W$. This consist of set $\{S_{i},S_{j}\}$, $i,j= 1,2,3,4$, 
 $i\neq j$.
    \item When three $S_{i}$ are in  non-general position in $W$. This consist of set $\{S_{i},S_{j},S_{k}\}$, $i,j,k=1,2,3,4$, $i \neq j \neq k$
    \item Finally, we have the leading singularity which is given when $S_{1}, S_{2}, S_{3}$ and $S_{4}$ meet at non-general position in $W$.
\end{enumerate}
The two-propagators case and the three-propagators case are similar to the analysis of Sections \ref{sec:bub} and \ref{sec:triangle}. 

As a demonstrative example of the two propagator case, we consider the case when $S_{1}$ and $S_{2}$ are in non-general position in $W$. Using Eq.\eqref{bubsethreholpseudo} we get the following two particle threshold and pseudo-threshold singularity 
\begin{align}
\text{Threshold:} \quad p_{2}^{2}= (m_{1}+ m_{2})^{2} \nonumber \\
 \text{Pseudo-threshold:} \quad  p_{2}^{2}= (m_{1}- m_{2})^{2}
\end{align}
In a similar way, we can consider other combinations of two propagators. There are a total of 6 such cases.

We can also consider the case when three propagators meet at non-general position. Consider the case when $S_{1}, S_{2}$ and $S_{3}$ are in  non-general position in $W$. For this case the singularity is given by Eq.\eqref{triangathresh} with the replacement $p_{3} \rightarrow p_{3}+p_{4} $. In a similar way, one can consider other combinations of the propagators and obtain the singularity with proper replacement in Eq.\eqref{triangathresh}. There are 4 such cases. 

Before proceeding we introduce a few variables so as to facilitate comparison with the literature. We use the following
\begin{align}
    s&=(p_{1}+p_2)^{2}, \; t=(p_{1}+p_3)^{2}, \nonumber \\
    p_1^2&=M_1^2, \; p_2^2=M_2^2, \; p_3^2=M_3^2, \; p_4^2=M_4^2, u=M_1^2+M_2^2+M_3^2+M_4^2-s-t 
\end{align}
We now consider the case when $S_{1}, S_{2}, S_{3}$ and $S_{4}$ are in non-general position in $W$. 
The first condition for non-general position (given in subsection \ref{appendix:ngpos})  gives
\begin{align}
    S_{i}=0, \quad i=1,2,3,4 \nonumber \\
    \sum_{i}^{5}x_{i}^{2}-x_{6}^{2}=0
\end{align}
Similarly, the second condition gives
\begin{align}
\alpha_{2}(2 p_2 +\alpha_{3}(2(p_2+p_3)) + \alpha_{4}(-2 p_1) +\alpha_{5} (2 x)&=0 \nonumber \\
   \alpha _1 \left(-m_1^2-1\right)+\alpha _2 \left(-m_2^2+p_2^2-1\right)+\alpha _3 \left(-m_3^2+t-1\right)-m_4^2+p_1^2+2 \alpha _5 x_5-1&=0 \nonumber \\\
   -2 \alpha _5+\alpha _1 \left(1-m_1^2\right)+\alpha _2 \left(-m_2^2+p_2^2+1\right)+\alpha _3 \left(-m_3^2+t+1\right)-m_4^2+p_1^2+1&=0
\end{align}
Doing the analysis as before we get the following condition for the singularity
\begin{align}\label{eq:boxsingfull}
&2 m_3^4 \left[-2 M_1^2 \left(M_2^2+s\right)+\left(M_2^2-s\right){}^2+M_1^4\right]+2 m_1^4 \left[-2 M_3^2 \left(M_4^2+s\right)+\left(M_4^2-s\right){}^2+M_3^4\right]+\nonumber \\
&2 m_4^4[-2 M_2^2(M_3^2+t)+(M_3^2-t){}^2+M_2^4]+2 m_2^4 \left[-2 M_1^2 \left(M_4^2+t\right)+\left(M_4^2-t\right){}^2+M_1^4\right] \nonumber \\
&2[-2 M_1^2 M_3^2 (M_2^2 M_4^2+s t)+(M_2^2 M_4^2-s t){}^2+M_1^4 M_3^4]+4m_4^2[M_1^2 M_3^2 (M_2^2-M_3^2+t)+\nonumber \\
&s t(M_3^2-t)+\left.M_2^2 \left(t \left(M_4^2+s\right)+M_3^2 \left(M_4^2-2 t\right)\right)-M_2^4 M_4^2\right]+4m_1^2\left[m_3^2 M_3^2 s+m_4^2 M_3^2 s\right.\nonumber \\
&-m_3^2 s^2+ m_2^2 M_4^2 s+m_3^2 M_4^2 s-m_4^2 M_3^4+m_2^2 M_4^2 M_3^2+m_4^2 M_4^2 M_3^2-2 M_4^2 M_3^2 s -m_2^2 M_4^4+m_2^2\nonumber \\
&(M_3^2+M_4^2-s)+t \left(m_2^2 \left(-M_3^2+M_4^2+s\right)+M_3^2 \left(m_4^2+s\right)+\left(m_4^2-s\right) \left(s-M_4^2\right)-2 m_3^2 s\right)+ \nonumber \\
& M_2^2 \left(m_3^2 \left(-M_3^2+M_4^2+s\right)+\left(m_4^{2}-M_4^{2}\right) \left(M_4^2-s\right)-2 m_2^2 M_4^2+M_3^2 \left(m_4^2+M_4^2\right)\right)+\nonumber \\
&M_1^2(m_3^2 (M_3^2-M_4^2+s)+ M_3^2 \left(-2 m_4^2-M_3^2+M_4^2+s\right)]-4m_3^2\left[M_1^4 M_3^2\right. +M_2^4 \left(m_4^2+M_4^2\right)+ \nonumber \\
& s \left(m_4^2 M_3^2+t \left(s-m_4^2\right)\right)-M_2^2 \left(m_4^2 \left(-2 M_4^2+s+t\right)+m_4^2 M_3^2+s \left(M_4^2+t\right)\right)-M_1^2\left(t (s-\right. \nonumber \\
&m_4^2)+ \left.\left.M_2^2 \left(m_4^2+M_3^2+M_4^2-2 s\right)+M_3^2 \left(m_4^2+s\right)\right)\right]-4m_2^2\left[M_1^4 \left(m_3^2+M_3^2\right)+t^2 \left(m_4^2+s\right)\right. + \nonumber \\
& M_4^2 \left(m_4^2 M_3^2-M_2^2 \left(m_3^2+m_4^2-M_4^2\right)+m_3^2 s\right)-t\left(m_4^2 M_3^2+m_3^2 s-2 m_4^2 s+M_4^2 s+m_4^2 M_4^2\right.+ \nonumber \\
& \left.M_2^2 \left(-m_3^2+m_4^2+M_4^2\right)\right)-M_1^2\left(t \left(m_4^2+M_3^2-2 M_4^2+s\right)+M_2^2 \left(m_3^2-m_4^2+M_4^2\right)\right. \nonumber \\
& \left.\left.m_3^2 \left(-2 M_3^2+M_4^2+s+t\right)++M_3^2 \left(m_4^2+M_4^2\right)\right)\right]=0
\end{align}
The above result for the leading singularity can also be obtained from the analysis of Landau equation and has been presented in \cite{eden2002analytic}, where further analysis related to the physicality of the above singularity has also been presented.
To further simplify the above result we make the substitution $M_{i}=M, m_{i}=m$, and get
\begin{align}
    2st \left(4 m^2 \left(-4 M^2+s+t\right)+4 M^4-s t\right)=0
\end{align}
We note that this result matches with the result given in \cite{Mizera:2021icv}, thus providing an important cross-check of Eq.\eqref{eq:boxsingfull}.

\section{Sunset integral}\label{sec:sunset}
\begin{figure}[H]
\centering\includegraphics[width=.35\textwidth]{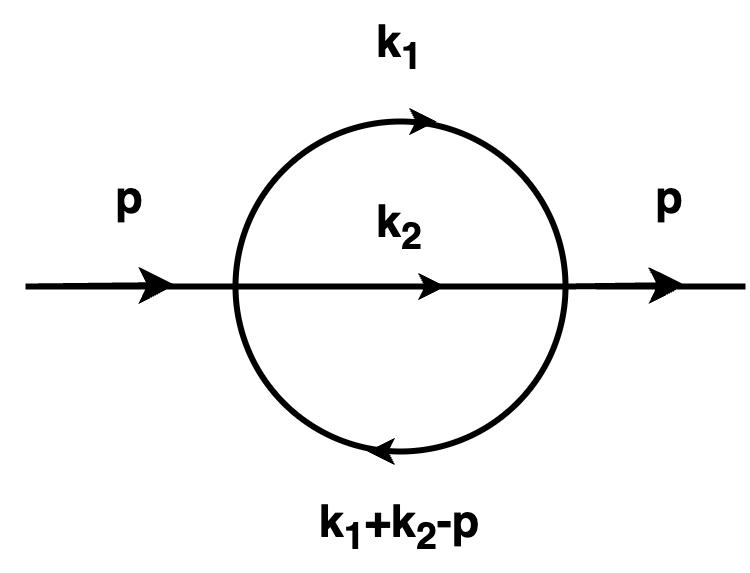}
\caption{Sunset diagram}\label{figsunset}
\end{figure}
We now consider the two-loop Sunset Integral as a starting point for the two-loop case. The sunset integral corresponding to the sunset diagram of Fig.\ref{figsunset} is given by 
\begin{equation}\label{sunsetint}
    I_{s}= \int \frac{d^{4}k_{1}d^{4}k_{2}}{(k_{1}^{2}-m_{1}^{2})((k_{2})^{2}-m_{2}^{2})((k_{1}+k_{2}-p)^2 - m_{3}^{2}))}
\end{equation}
The three propagators are as follows: \\
\begin{center}
 $k_1^2-m_1^2, \quad k_2^2-m_2^2$ and $(k_1 +k_2-p)^2 -m_3^2$ \\   
\end{center}

For the first two propagators, the compactification procedure is similar to the one-loop case with different variables. The compactification for the third propagator is non-trivial and has to be done recursively, as has been outlined in Appendix \ref{appendix:compact}. After compactification, we get the following compactified propagators
\begin{align}
    S_1 & = \frac{x_5 (-m_1^2 - 1) + x_6 (-m_1^2 + 1)}{x_{5}+x_{6}}\nonumber \\
    S_2 &= \frac{y_5 (-m_2^2 - 1) + y_6 (-m_2^2 + 1)}{{y_{5}+y_{6}}}\nonumber \\
    S_3 & = ((y_5 + y_6) (x_5 (-m_3^2 - 1) + x_6 (-m_3^2 + 1)) + (-2 p.y + 
     y_5 (p^2 - 1) + y_6 (p^2 + 1)) (x_5 + x_6)\nonumber \\ 
     &+ 2 x.y - 2 p.x (y_5 + y_6))\frac{1}{(x_{5}+x_{6})(y_{5}+y_{6})}
\end{align}
It is to be noted that we can write $S_{3}$ in a more suggestive manner as 
\begin{equation}
  S_3  = p^{2} -m_{3}^{2} +\frac{x_{6}-x_{5}}{x_{5}+x_{6}} +\frac{y_{6}-y_{5}}{y_{5}+y_{6}} - \frac{2 p \cdot y}{y_{5}+y_{6}} -\frac{2 p \cdot x}{x_{5}+x_{6}}        
\end{equation}
We again remark that the above propagator is homogeneous in $\mathbf{x}$ and $\mathbf{y}$.

The new ambient space is given by 
$$W_{1} \times W_{2} \subset \mathbb{CP}^{5} \times \mathbb{CP}^{5}$$
where 
\begin{align}
    W_1&= \{(x_{1},\cdots,x_{6})|\sum_{i}^{5}x_{i}^{2}-x_{6}^{2}=0 \}\subset \mathbb{CP}^5 \nonumber \\
    W_2&= \{(y_{1},\cdots,y_{6})|\sum_{i}^{5}y_{i}^{2}-y_{6}^{2}=0 \}\subset \mathbb{CP}^5
\end{align}
The analysis is the same as in the previous section, though more tedious due to a large number of equations arising from the conditions of meeting at non-general position. We will focus only on the leading singularity for the present case as it is the non-trivial one. Other singularities can be obtained using the result of previous sections.

The leading singularity occurs when $S_1, S_2 $ and $S_3$ meet at non-general position in $W_1 \times W_2$. The analysis for this case is tedious and has been done using \textsc{Mathematica}. We outline the important steps of the calculation. Using the first condition for the hyper-planes to meet at non-general position, we get 
\begin{align}
    S_{i}= 0, i=1,2,3 \nonumber \\
    \sum_{i}^{5}x_{i}^{2}-x_{6}^{2}=0, \quad \sum_{i}^{5}y_{i}^{2}-y_{6}^{2}=0
\end{align}
Using the second condition we get 
\begin{align}
\alpha_{3}( (2x) -2 p(y_{5}+y_{6})) + \alpha_{4}(2 y) &=0 \nonumber \\
\alpha_{2}(-m_{3}^{2}-1) +\alpha_{3}((y_{5}+y_{6})(-m_{3}^{2}-1) +(-2 y\cdot p + y_{5}(p^{2}-1)+y_{6}(p^{2}+1))+\alpha_{4}(2x_{5}) &=0 \nonumber \\
\alpha_{2}(-m_{3}^{2}+1) +\alpha_{3}((y_{5}+y_{6})(-m_{3}^{2}+1) +(-2 y\cdot p + y_{5}(p^{2}-1)+y_{6}(p^{2}+1))+\alpha_{4}(2x_{6}) &=0 \nonumber \\
\alpha_{3}((2 y) -2 p(y_{5}+y_{6}))+\alpha_{4}(2x) &=0 \nonumber \\
\alpha_{3}(-m_{2}^2-1)+\alpha_{3}(x_{5}(-m_{3}-1)+x_{6}(-m_{3}^{2}+1)+(p^{2}-1)(x_{5}+x_{6})-2 p\cdot x)+\alpha_{5}(2y_{5})&=0  \nonumber \\
\alpha_{3}(-m_{2}^2+1)+\alpha_{3}(x_{5}(-m_{3}^{2}-1)+x_{6}(-m_{3}^{2}+1)+(p^{2}-1)(x_{5}+x_{6})-2 p\cdot x)+\alpha_{5}(2y_{5})&=0 
\end{align}
Repeating the analysis as in Section \ref{sec:bub} and simplifying, we get the following singularities corresponding to the Sunset integral 
\begin{align}
   &p^2 = (-m_1 + m_2 + m_3)^2, \; p^2 = (m_1 + m_2 - m_3)^2, \; p^2 = (m_1 - m_2 + 
   m_3)^2, \nonumber \\
   & p^2 = (m_1 + m_2 + m_3)^2
\end{align}
The first three singularities are called the pseudo-threshold singularities and the last singularity is called the threshold singularity \cite{Berends:1997vk}.

\section{Double box integral}\label{sec:dbox}
\begin{figure}[H]
\centering\includegraphics[width=.35\textwidth]{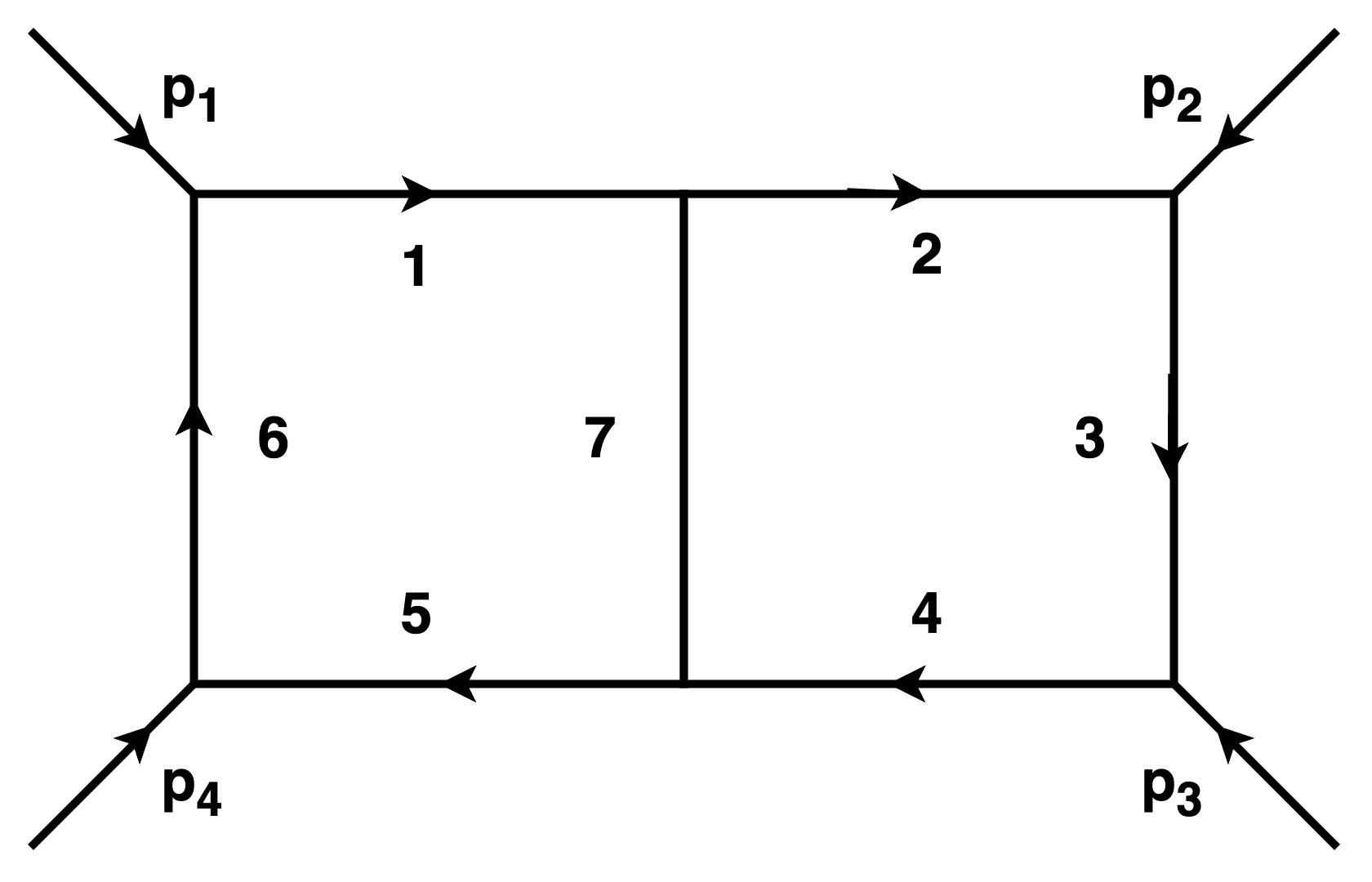}
\caption{Sunset diagram}\label{figdoublebox}
\end{figure}
Next, we consider the case of Double Box integral corresponding to the double box diagram in Fig.\ref{figdoublebox}
\begin{align}\label{dboxintegral}
I_{4,2}= \int \int &\frac{d^{4}k_{1}d^{4}k_{2}}{(k_{1}^{2}-m_{1}^{2})(k_{2}^{2}-m_{2}^{2})((k_{2}+p_{2})^{2}-m_{3}^{2})((k_{2}+p_{2}+p_{3})^{2}-m_{4}^{2})}\nonumber \\
\times &\frac{1}{((k_{1}+ p_{2}+p_{3})^{2}-m_{5}^{2})((k_{1}+ p_{2}+p_{3}+p_{4})^{2}-m_{6}^{2})((k_{1}-k_{2})^{2}-m_{7}^{2})}
\end{align}
The compactified propagators are as follows
\begin{align}
    S_1 &= \frac{\left(-m_{1}^2-1\right) x_5+\left(1-m_{1}^2\right) x_6}{x_{5}+x_{6}} \nonumber \\
    S_2&=\frac{\left(-m_{2}^2-1\right) y_5+\left(1-m_{2}^2\right) y_6}{x_{5}+x_{6}} \nonumber \\
    S_3&=\frac{y_5 \left(-m_{3}^2+p_2^2-1\right)+y_6 \left(-m_{3}^2+p_2^2+1\right)+2 p_2.y}{x_{5}+x_{6}} \nonumber \\
    S_4&= \frac{y_5 \left(-m_{4}^2+t-1\right)+y_6 \left(-m_{4}^2+t+1\right)+2 \left(p_2.y+p_3.y\right)}{x_{5}+x_{6}}\nonumber \\
    S_5&= \frac{x_5 \left(-m_{5}^2+t-1\right)+x_6 \left(-m_{5}^2+t+1\right)+2 \left(p_2.x+p_3.x\right)}{x_{5}+x_{6}}\nonumber \\
    S_6&=\frac{x_5 \left(-m_{6}^2+p_{1}^2-1\right)+x_6 \left(-m_{6}^2+p_{1}^2+1\right)-2 p_{1}.x}{x_{5}+x_{6}} \nonumber \\
    S_7&=\frac{\left(\left(-m_{7}^2-1\right) x_5+\left(1-m_{7}^2\right) x_6\right) \left(y_6+y_5\right)-2 x.y+\left(x_5+x_6\right) \left(y_6-y_5\right)}{(x_{5}+x_{6})(y_{5}+y_{6})}
\end{align}

To demonstrate the method, we consider a few cases where the results of the sunset integral can be used. As an example consider the case when the  surfaces $\{ S_{i},S_{j},S_{7}\}$, $i=1,5,6$ and $j=2,3,4$, are in non-general position in $W$. Then we have the following singularities
\begin{itemize}
    \item $\{ S_{1},S_{2},S_{7}\}$ : $(p_{1}+p_{2})^{2}= (m_{1}+m_{2}+m_{7})^{2},(m_{1}-m_{2}+m_{7})^{2},(m_{1}+m_{2}-m_{7})^{2}, (-m_{1}+m_{2}+m_{7})^{2}$
    \item $\{ S_{1},S_{3},S_{7}\}$ : $(p_{1}+p_{3})^{2}= (m_{1}+m_{3}+m_{7})^{2},(m_{1}-m_{3}+m_{7})^{2},(m_{1}+m_{3}-m_{7})^{2}, (-m_{1}+m_{3}+m_{7})^{2}$
    \item $\{ S_{1},S_{4},S_{7}\}$ : $(p_{1}+p_{4})^{2}= (m_{1}+m_{4}+m_{7})^{2},(m_{1}-m_{4}+m_{7})^{2},(m_{1}+m_{4}-m_{7})^{2}, (-m_{1}+m_{4}+m_{7})^{2}$
    \item $\{ S_{5},S_{2},S_{7}\}$ : $(p_{5}+p_{2})^{2}= (m_{5}+m_{2}+m_{7})^{2},(m_{5}-m_{2}+m_{7})^{2},(m_{5}+m_{2}-m_{7})^{2}, (-m_{5}+m_{2}+m_{7})^{2}$
    \item $\{ S_{5},S_{3},S_{7}\}$ : $(p_{5}+p_{3})^{2}= (m_{5}+m_{3}+m_{7})^{2},(m_{5}-m_{3}+m_{7})^{2},(m_{5}+m_{3}-m_{7})^{2}, (-m_{5}+m_{3}+m_{7})^{2}$
    \item $\{ S_{5},S_{4},S_{7}\}$ : $(p_{5}+p_{4})^{2}= (m_{5}+m_{4}+m_{7})^{2},(m_{5}-m_{4}+m_{7})^{2},(m_{5}+m_{4}-m_{7})^{2}, (-m_{5}+m_{4}+m_{7})^{2}$
    \item $\{ S_{6},S_{2},S_{7}\}$ : $(p_{6}+p_{2})^{2}= (m_{6}+m_{2}+m_{7})^{2},(m_{6}-m_{2}+m_{7})^{2},(m_{6}+m_{2}-m_{7})^{2}, (-m_{6}+m_{2}+m_{7})^{2}$
    \item $\{ S_{6},S_{3},S_{7}\}$ : $(p_{6}+p_{3})^{2}= (m_{6}+m_{3}+m_{7})^{2},(m_{6}-m_{3}+m_{7})^{2},(m_{6}+m_{3}-m_{7})^{2}, (-m_{6}+m_{3}+m_{7})^{2}$
    \item $\{ S_{6},S_{4},S_{7}\}$ : $(p_{6}+p_{4})^{2}= (m_{6}+m_{4}+m_{7})^{2},(m_{6}-m_{4}+m_{7})^{2},(m_{6}+m_{4}-m_{7})^{2}, (-m_{1}+m_{2}+m_{7})^{2}$
\end{itemize}

We remark that the above result can also be obtained using the Landau equation analysis as presented in \cite{eden2002analytic}.
Next, we consider the second type singularities. The mechanism for the second type singularities to occur is different from the previous cases as they arise due to the singular manifold $S_{7}$ in the present case \cite{federbush1965note}. These singularity occur because in momentum space $S_{7}$ which is given by $(k_{1}-k_{2})^{2}-m_{7}^{2}$, corresponds to line $k_{1}= k_{2} \pm m_{7}$ which intersect at infinity in the $k_{1},k_{2} $ plane.

The second type singularities occur when $\{S_{i},S_{j}\}$, $i,j= 1,2,3,4,5,6, i\neq j$, meet in non-general position \cite{federbush1965note}
\begin{itemize}
    \item $\{S_{6},S_{2}\}$ : $p_{1}^{2} = 0$
    \item $\{S_{6},S_{3}\}$ : $(p_{1}+p_{2})^{2} = 0$
\end{itemize}
Similarly, we can also obtain other second-type singularities for other combinations
with proper substitutions of $i$ and $j$.

We can also consider a case with 3 propagators as follows
\begin{align*}
    \{S_{6},S_{1},S_{3}\}: p_{1}^{2}p_{2}^{2}= (p_{1}\cdot p_{2})^{2}
\end{align*}
In a similar we can obtain other singularities with proper substitutions of $i$ and $j$.

\section{Summary and discussion}
We considered one and two-loop Feynman integrals and studied the singularities associated with them using the method extending the analysis in \cite{hwa1966homology}. We found for the tractable cases of one loop Bubble and the Triangle integrals it is possible to determine whether the singularity lies on the physical sheet or not. We found parallels with the properties of Feynman parameters for singularities in physical sheet \cite{Coleman:1965xm}. The analysis of the second type of singularity was presented for both one-loop cases, where they occur due to the presence of an effective denominator, and for the two-loop cases where a different mechanism is responsible for them \cite{federbush1965note}. We showed that by extending the analysis presented in \cite{federbush1965note} such a technique can also be used to obtain singularities of other kinds. Thus, the results presented here in our opinion, constitute important advances in our knowledge of the structure of Feynman integrals, which are the basic building blocks of perturbative quantum field theory, on which our entire knowledge of the standard model rests. By bringing in methods from algebraic geometry and applying them to the concrete problem of Landau and non-landau singularities, we have, in our opinion provided insights into their singularity structure,  thereby exploring a new frontier in fundamental physics that rests on mathematics, and is independent of whether the interactions arise from the SM or beyond. 

We remark that the calculation becomes tedious as the number of propagators increases and thus the procedure asks for proper optimization and automation. There are other works that can be done in connection with the present analysis. The analysis of the two-loop case is not complete due to several technical difficulties. Another important direction which was not presented in the current analysis is the construction of the Kronecker index table \cite{hwa1966homology} to determine the `full sheet structure' of these integrals. This table would give us the knowledge of the sheet structure in an algebraic manner as has been shown for a simple unitary integral in \cite{hwa1966homology}. The construction of the Kronecker index table requires the construction of vanishing cycles. The way to construct such vanishing cycles for various one loop examples has been outlined in \cite{boyling1966construction} and \cite{boyling1967homological}. The Kronecker table also allows us to apply Picard–Lefschetz theorem\cite{hwa1966homology,Bogner:2017vim}, which is further crucial in determining the full sheet structure of these integrals.

A further application of the Picard–Lefschetz theorem is the calculation of the discontinuity around a singularity. In \cite{hwa1966homology}, a generalized version of Cutkosky's discontinuity formula is discussed along with an example of unitarity integral which can be extended to cases shown in this paper. Another important analysis is the calculation of the homology group related to these Feynman integrals. For the one loop cases , it is called the decompositon theorem and is presented in \cite{hwa1966homology}. For the two-loop cases, a detailed analysis using Double Box integral as example, but without momentum conservation is presented in \cite{federbush1965calculation}. Similar analysis related to the computation of homology groups, motivated by, and for further use in, Feynman integrals has also been studied in \cite{muhlbauer2022homology}. We would also like to mention the recent work \cite{muhlbauer2020momentum,muhlbauer2022cutkosky}, where analysis of Landau equations and a proof of Cutkosky's theorem for massive Feynman integrals were presented using related techniques used here. Thus, homological methods provide a universal framework to study these properties of Feynman integrals.

\section{Acknowledgments}
The authors would like to thank B. Ananthanarayan for proposing the current investigation and for providing useful comments. The authors would also like to thank Souvik Bera, and Sudeepan Datta for their contribution during the initial stages of the project and also the Centre for High Energy Physics, Indian Institute of Science Bangalore, where this work was done. This work is a part of TP's doctoral work at CHEP, IISc.
\appendix
\section{Appendix: A toy example}\label{appendix:toy}
\begin{figure}[htbp]
\centering\includegraphics[width=.4\textwidth]{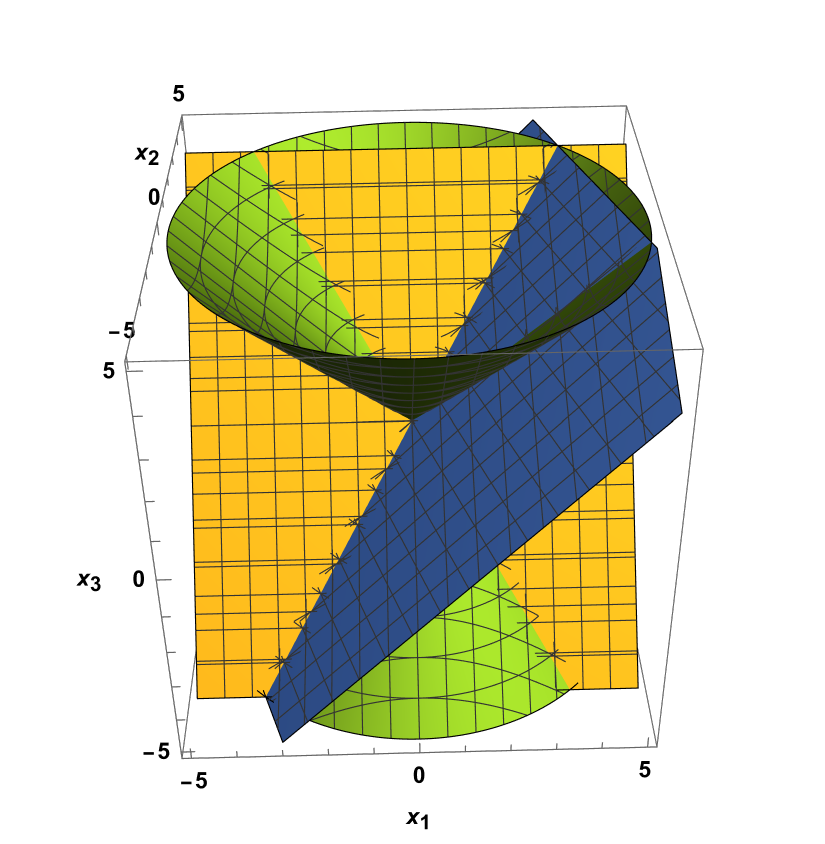}
\caption{$S_{1}$ and $S_{2}$ (in blue and yellow colour respectively) meeting in non-general position in $W$(in green).}\label{fig:bubngpos3d}
\end{figure}

\begin{figure}[H]
		\centering
		\begin{subfigure}[b]{0.4\textwidth}
		\centering
		\includegraphics[width=.73\textwidth]{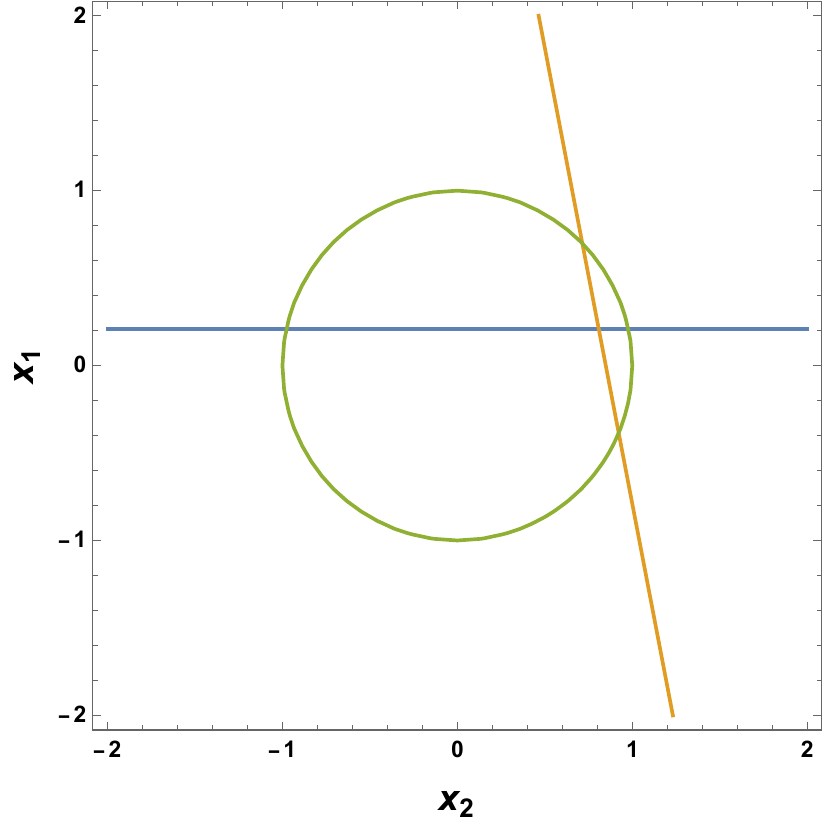}
			\caption{}
			\label{fig:generalposbub2d}
		\end{subfigure}
		\hfill
		\begin{subfigure}[b]{0.42\textwidth}
		\centering
		\includegraphics[width=.7\textwidth]{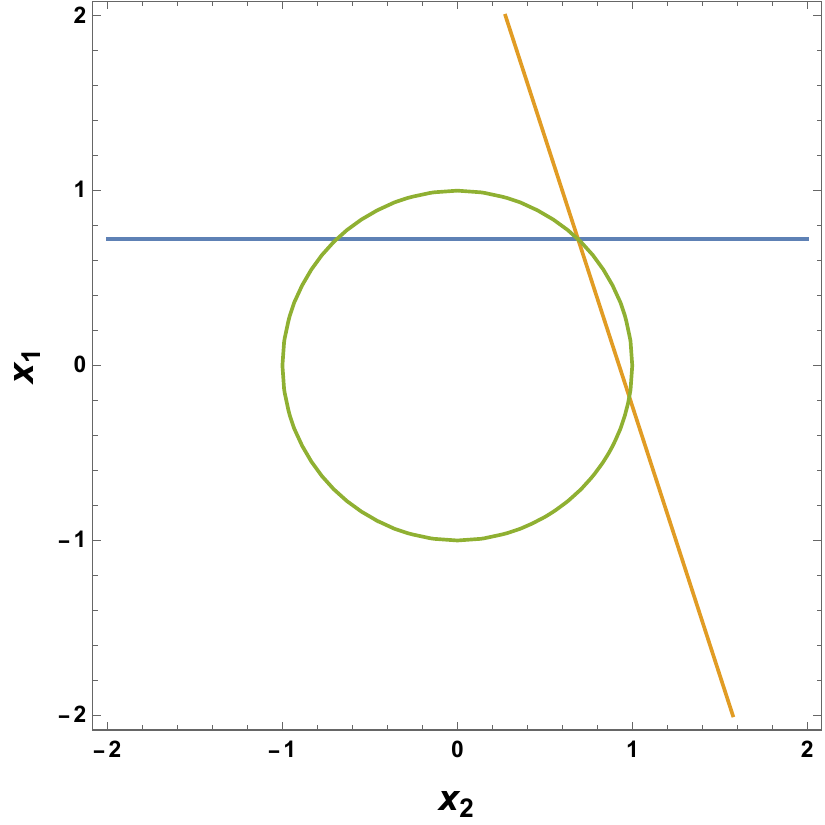}
			\caption{}
			\label{fig:nongeneralposbub2d}
		\end{subfigure}
		\caption{(a) $S_{1}$ and $S_{2}$ meeting in general position in $W$. (b) $S_{1}$ and $S_{2}$ meeting in non-general position in $W$, corresponding to singularity $p^{2}=(m_{1}+m_{2})^{2}$. The plots shows the situation shown in Fig.\ref{fig:bubngpos3d} in $(x_{1},x_{2})$-plane with $x_{3}=1$.  }
		\label{2dplots}
\end{figure}

In this appendix, we consider a toy example to demonstrate the method. We consider the following one-dimensional version of the bubble integral
\begin{equation}\label{inttoybub}
    I_{2} = \int_{\mathbb{R}} \frac{dk}{(k^{2}-m_{1}^{2})((k-p)^{2}-m_{2}^{2})}
\end{equation}
Here is the integration cycle is $\mathbb{R}$ and the ambient space is $\mathbb{C}$, so we use the compactification procedure outlined in  Section \ref{appendix:compact}.
Compactifying the propagators we get the following
\begin{align}
    S_1 &= \frac{x_{2}(-m_{1}^{2}-1)+ x_{3}(-m_{2}^{2}+1)}{x_{2}+x_3}\nonumber \\ 
    S_2&= \frac{-2 p x_{1} +x_{2}(p^{2}-m_{2}^{2}-1)+x_{3}(p^{2}-m_{1}^{2}+1)}{x_{2}+x_3}
\end{align}
and the new ambient space $W$ is given by 
\begin{equation*}
    W= \{(x_{1},x_{2},x_{3})|x_{1}^{2}+x_{2}^{2}-x_{3}^{2}=0 \}\subset \mathbb{CP}^2
\end{equation*}
Similarly, we have $dk \rightarrow \frac{dx_{1}}{x_{2}+x_{3}}$. This gave rise to an effective denominator 
\begin{equation}
    S_{3}= x_{2}+x_{3}
\end{equation}
The singularities of integral \eqref{inttoybub}  correspond to the following two cases
\begin{enumerate}
    \item When $S_{1}$ and $S_{2}$ meet in non-general position in $W$. This case is similar to the case of the Bubble Integral in Section\ref{sec:bub}. We get the two singularities:  $p^{2}= (m_{1}+m_{2})^{2}$ and $p^{2}=(m_{1}-m_{2})^2$. 
    \item When $S_{1},S_{2}$ and $S_{3}$ meet in non-general position in $W$. This case is similar to the Bubble Integral case and we get $p^{2}=0$.
\end{enumerate}
The situation for the $S_{1}$ and $S_{2}$ meeting in non-general position (for real $x_{1}, x_{2}, x_{3}$) is as shown in Fig. \ref{fig:bubngpos3d}.
We can also look at the situation in the real $(x_{1},x_{2})$-plane (with $x_{3}=1$). The situation of general and non-general positions is shown in Fig. \ref{2dplots}.
We note a special feature of the intersection of these surfaces is the `vanishing cycle'. In the plots shown in Fig. \ref{2dplots},  notice that the green circle is divided into four parts in Fig. \ref{fig:generalposbub2d} and into three parts in Fig.\ref{fig:nongeneralposbub2d}. The region that vanishes due to the meeting of the surfaces at non-general position is called the `vanishing cycle'. So whenever the surfaces meet at non-general position it corresponds to the vanishing of a cycle.


\section{Conflict of Interest Statement}
We have no conflicts of interest to disclose.
\section{Data Availability Statement}
Data sharing is not applicable to this article as no datasets were generated or analysed during the current study.

\bibliographystyle{JHEP}
\bibliography{biblio} 

\end{document}